\documentclass[aps,pre,showpacs,amsmath,amssymb,amsfonts,superscriptaddress,lengthcheck,longbibliography]{revtex4-2}

\usepackage{graphicx}\graphicspath{ {figures/} }
\usepackage{hyperref}
\hypersetup{colorlinks,allcolors=blue,breaklinks}

\newcommand{\be}{\begin{equation}}
\newcommand{\ee}{\end{equation}}
\newcommand{\ba}{\begin{align}}
\newcommand{\ea}{\end{align}}
\newcommand{\bi}{\begin{itemize}}
\newcommand{\ei}{\end{itemize}}

\newcommand{\bla}{bla\\bla\\bla\\bla\\bla}

\begin{document}

\title{A self-consistent criterion for the range of validity of weakly driven processes}

\author{Pierre Naz\'e}
\email{pnaze@ufpa.br}

\affiliation{\it Universidade Federal do Par\'a, ICEN, Faculdade de F\'isica, 
Av. Augusto Corr\^ea, 1, Guam\'a, 66075-110, Bel\'em, Par\'a, Brazil}

\date{\today}

\begin{abstract}
One of the longstanding open questions in linear response theory concerns its true range of validity. Determining when the linear approximation can be trusted typically requires knowledge of second-order corrections, which are often difficult to compute explicitly. In this letter, I propose a self-consistent criterion for the validity of linear response, formulated in terms of a typical length scale that emerges from the fluctuation-response inequality within the theory itself. The result applies to classical open systems. I illustrate the criterion with explicit examples of Brownian particles in harmonic traps, and classical open systems presenting Kibble-Zurek mechanism. Finally, I discuss the physical meaning of this typical length, providing both thermodynamic and information-theoretic interpretations.
\end{abstract}

\maketitle

\section{Introduction}

Linear response theory is one of the central pillars of nonequilibrium statistical mechanics~\cite{kubo1957statistical}. It provides a remarkably powerful framework: when a system is weakly perturbed from equilibrium, its response can be expressed entirely in terms of equilibrium correlation functions. From transport coefficients to fluctuation–dissipation relations, linear response connects microscopic dynamics to measurable macroscopic behavior in a way that is both elegant and practical.

Yet, despite its wide use and conceptual clarity, a fundamental question remains surprisingly subtle: how weak is ``weak'' driving? In practice, the condition is often stated heuristically: one assumes that the change in the control parameter is small compared to its initial value~\cite{kubo1957statistical}. However, this assumption is rarely examined in a systematic way. The actual range of validity of linear response theory depends not only on the amplitude of the perturbation, but also on intrinsic properties of the system and its coupling to the environment. Establishing a quantitative and physically meaningful criterion for this range of validity is therefore an important open problem.

In this work, I propose a self-consistent criterion for weakly driven processes based on the fluctuation-response inequality (FRI)~\cite{dechant2020FRI,ItoDechant2020,kwon2025fluctuation,dechant2025finite,wang2020sub,zheng2025unified,aslyamov2025nonequilibrium,chun2026fluctuation,owen2020universal,Shiraishi_2023}. Rather than estimating higher-order corrections directly, I exploit the structure of the relative entropy between the nonequilibrium and equilibrium states. This allows us to identify a typical thermodynamic length scale, determined by equilibrium fluctuations, that naturally bounds the admissible driving strength. The condition that emerges is independent of the particular driving protocol and depends only on equilibrium properties of the system.
In open systems, the relative entropy acquires a clear thermodynamic interpretation in terms of the absorbed heat of the system during the driving.

To illustrate the criterion, I analyze overdamped Brownian motion in harmonic traps, considering both moving and stiffening protocols~\cite{naze2022optimal}. These examples corroborate the typical parameter ranges commonly assumed in the literature, but now grounded in a principled inequality. I also discuss the case of classical open systems presenting the Kibble-Zurek mechanism, which leads to a failure of linear response theory at criticality. Finally, we discuss thermodynamic and information-geometric interpretations of the result, showing that the condition for linear response can be understood in terms of Fisher information~\cite{LyMarsmanWagenmakers2017,Zegers2015FisherIP,BerishaHero2015EmpiricalFI,Wang2024FisherRSS,Verkuilen2022FisherFunc,Kharazmi2022JensenFisher,Nielsen2007FisherChiDistances,SanchezMoreno2012JensenFisher,BalakrishnanStepanov2006FisherRecord,MeloQueirosMorgado2025}.

In this way, the present letter reframes the question of linear response validity: instead of a heuristic smallness assumption, we obtain a system-dependent bound derived from fundamental fluctuation relations. To the best of our knowledge, this explicit, system-dependent bound on the admissible driving strength has not been isolated in this form.

\section{Weakly driven processes}
\label{sec:wdp}

Consider a classical system with a time-dependent Hamiltonian $\mathcal{H}(\lambda(t))$, where $\lambda(t)$ is the external parameter that drives the system. Initially, the system is prepared at thermal equilibrium with a weakly coupled thermal reservoir with temperature $\beta^{-1}$. During the driving, the system is open, meaning that the heat bath is in contact with it. I also consider that the external parameter is
\be
\lambda(t)=\lambda_0+g(t/\tau)\delta\lambda,
\ee
where $\lambda_0$ is its initial value, $\delta\lambda$ is its driving strength, $g(t/\tau)$ is the protocol, and $\tau$ the switching time of the process. In particular, the driving is weak, where I assume a perturbative regime in which the probability distribution can be expanded in powers of $\delta\lambda$. To measure the thermodynamic quantities, such as averaged thermodynamic work, one uses the relaxation function~\cite{kubo1957statistical}
\be
\Psi_0(t) = \beta\langle\overline{\partial_\lambda\mathcal{H}(t)} \partial_\lambda\mathcal{H}(0)\rangle_0-\langle\partial_\lambda\mathcal{H}\rangle_0^2,
\label{eq:relaxfunc}
\ee
where $\langle\cdot\rangle_0$ is the average at the initial canonical ensemble and $\overline{\cdot}$ the average over the noise produced by the heat bath along the driving. I assume that the relaxation function is symmetric in time, $\Psi_0(t)=\Psi_0(-t)$, with a positive Fourier transform, in order to satisfy the Second Law of Thermodynamics~\cite{naze2020compatibility}. Here, $\Psi_0(0)$ is the equilibrium variation of the generalized force $\partial_\lambda\mathcal{H}$.

The objective of this letter is to derive the thermodynamic uncertainty relation in weakly driven processes, and use such inequality to deduce a typical length $\ell_0$ that will intermediate the range of validity of linear response theory. I will present examples using overdamped Brownian motions in harmonic traps, corroborating the range of validity already employed in previous works. Interpretations of its thermodynamic or informational aspects are also discussed.

It is important to stress that the derivation assumes the validity of linear response in the Kubo sense and the differentiability of the equilibrium family of states with respect to the control parameter. The bound derived below should therefore be interpreted as a self-consistency condition within this perturbative framework: it identifies the intrinsic amplitude scale beyond which the linear approximation ceases to remain controlled. In this sense, the result does not presuppose smallness, but quantifies it.

\section{Fluctuation-response inequality}
\label{sec:tur}

Consider $\rho(\Gamma,t)$ the non-equilibrium probability distribution expanded up to second order. It is called $D_{\rm KL}(\rho(t)||\rho_0)$ the relative entropy between $\rho(\Gamma,t)$ and the initial thermal probability distribution $\rho_0(\Gamma)$ (see Appendix~\ref{app:A}). Consider now $B(\Gamma)$ as an observable. For a thermally isolated and open system, and any type of observables, it is possible to show that holds the FRI (see Appendix~\ref{app:B})
\be
\frac{\langle B^2\rangle_0-\langle B\rangle_0^2}{(\langle B\rangle_t-\langle B\rangle_0)^2}\ge \frac{1}{\delta\lambda^2\int_{\Gamma} \frac{\rho_1^2(\Gamma',t)}{\rho_0(\Gamma')}d\Gamma'},
\label{eq:tur0}
\ee
where $\langle\cdot\rangle_t$ is the average in the non-equilibrium probability distribution, and the left-hand side of Eq.~\eqref{eq:tur0} quantifies the ratio between equilibrium fluctuations and the squared induced response. When it is small, that is, with a high response correction, it is necessary to have a high relative entropy between the probability distributions. In particular, for open systems, this manifests as a high absorption of heat by the system. Such a relation has been discovered and explored already in several contexts~\cite{dechant2020FRI,ItoDechant2020,kwon2025fluctuation,dechant2025finite,wang2020sub,zheng2025unified,aslyamov2025nonequilibrium,chun2026fluctuation,owen2020universal,Shiraishi_2023}.

\subsection{Quasistatic limit}

In the quasistatic limit, the relative entropy expanded until its second order is given by its quasistatic value (see Appendix~\ref{app:c})
\be
\int_{\Gamma} \frac{\rho_1^2(\Gamma',t)}{\rho_0(\Gamma')}d\Gamma'=\beta \Psi_0(0).
\label{eq:heatqs}
\ee
Under which conditions the FRI becomes an equality? Observe that the precision will depend on the observable $\partial_\lambda\mathcal{H}$ via $\Psi_0(0)$ if the equality exists. In particular, choosing $B=\partial_\lambda\mathcal{H}$, one has in the quasistatic limit
\be
\rho_1 = \frac{d\rho_0}{d\lambda} \propto (\partial_\lambda \mathcal{H}-\langle\partial_\lambda\mathcal{H}\rangle_0) \cdot \rho_0,
\ee
which is the necessary and sufficient condition for the Cauchy-Schwarz inequality to become an equality. In this manner
\be
\frac{\langle \partial_\lambda\mathcal{H}^2\rangle_0-\langle \partial_\lambda\mathcal{H}\rangle_0^2}{(\langle \partial_\lambda\mathcal{H}\rangle_{\rm qs}-\langle \partial_\lambda\mathcal{H}\rangle_{0})^2}= \frac{1}{\beta \Psi_0(0)\delta\lambda^2},
\label{eq:tur}
\ee
with $(\langle \partial_\lambda\mathcal{H}\rangle_{\rm qs}-\langle \partial_\lambda\mathcal{H}\rangle_{0})^2\neq 0$. Now enter the key point of this letter. Since $\langle \partial_\lambda\mathcal{H}^2\rangle_0-\langle \partial_\lambda\mathcal{H}\rangle_0^2$ counts the zeroth order ($\mathcal{O}(1)$), while $( \langle \partial_\lambda\mathcal{H}\rangle_{\rm qs}-\langle \partial_\lambda\mathcal{H}\rangle_{0})^2$ not ($\mathcal{O}(\delta\lambda^2)$). Then $\langle \partial_\lambda\mathcal{H}^2\rangle_0-\langle \partial_\lambda\mathcal{H}\rangle_0^2\gg (\langle \partial_\lambda\mathcal{H}\rangle_{\rm qs}-\langle \partial_\lambda\mathcal{H}\rangle_{0})^2$. Indeed, for linear response to remain controlled, the response correction must remain small compared to equilibrium fluctuations. One can show now
\be
\delta\lambda\ll \frac{1}{\sqrt{\beta \Psi_0(0)}}:=\ell_0.
\label{eq:condlrt}
\ee
Calling $\ell_0$ the typical length, the driving strength $\delta\lambda$ always needs to be much less than its value for linear response theory to work out. In other words, to have thermodynamic quantities well predicted using the framework of weakly driven processes. Additionally, the result does not depend on the driving process, but only on the equilibrium properties of the system and bath.

Also, the relation of the typical length can be rewritten in the following form
\be
\Psi_0(0)\,\delta\lambda^2\ll k_B T,
\ee
meaning that linear response theory holds when the equilibrium variation of the generalized force multiplied by the squared perturbation, {\it i.e.}, the equilibrium perturbation of the system energy due to the driving, is small compared to the equilibrium thermal fluctuations. This is a natural result to expect.

Right now we have a new way to measure the range of validity of linear response theory, by using
\be
\frac{\delta\lambda}{\ell_0}\ll 1,
\label{eq:main}
\ee
which is the main result of this letter. This condition emerges then as a self-consistency requirement: it guarantees that response corrections remain parametrically smaller than zeroth-order fluctuations, ensuring that higher-order contributions do not dominate. The result is thus not a tautological restatement of perturbation theory, but an intrinsic amplitude scale --- determined entirely by equilibrium properties --- that must be respected for the perturbative expansion to remain controlled. To obtain better results in such a relation, tighter FRIs are necessary to improve $\ell_0$. This work is just the first step.

In general, examples treated in the literature use unit values, such that $\beta=1,\Psi_0(0)=1,\lambda_0=1$. Thus, the condition~\eqref{eq:main} becomes
\be
\frac{\delta\lambda}{\lambda_0}\ll 1,
\ee
which is the basic condition required in the hypothesis of linear response theory in some articles~\cite{naze2020compatibility,naze2022optimal}. Therefore, the procedure is logically consistent as it is seen in the agreements in the examples of those works. 

It is important to stress that the bound obtained here is not merely a statement about the smallness of the Kullback–Leibler divergence between nearby equilibrium states. Rather, it arises from a FRI that directly constrains the observable response itself. The inequality links the first-order correction of expectation values to equilibrium fluctuations through a universal Cauchy–Schwarz structure. In this sense, the condition $\delta\lambda\ll \ell_0$ ensures that the response of any observable remains subordinate to intrinsic equilibrium fluctuations. The bound therefore operates at the level of measurable quantities, not only at the level of distributional distinguishability. Let us see some examples now.

\section{Examples}

\subsection{Moving trap: degenerate case}

Consider the white noise overdamped Brownian motion, subject to a moving trap $\lambda(t)$, and unit mass~\cite{naze2022optimal}. Observe that
\be
\langle \partial_\lambda\mathcal{H}\rangle_{\rm qs}-\langle \partial_\lambda\mathcal{H}\rangle_{0}= 0.
\ee
Therefore, our typical length does not apply here. Remind the reader that such an example is exact in linear response theory since its dynamics is linear~\cite{naze2022optimal}. Therefore, a regime of weak perturbation is pointless. Using relation~\eqref{eq:tur0}, one can show
\be
\delta\lambda^2\ge 0.
\ee
Justifying that every perturbation $\delta\lambda$ can be applied. Indeed, the moving trap is a degenerate case.

\subsection{Stiffening trap: non-degenerate case}

Consider a white noise overdamped Brownian motion, subject to a stiffening trap $\lambda(t)$, and unit mass~\cite{naze2022optimal}. Observe that
\be
(\langle \partial_\lambda\mathcal{H}\rangle_{\rm qs}-\langle \partial_\lambda\mathcal{H}\rangle_{0})^2= \frac{\delta\lambda^2}{4\beta^2 \lambda_0^4}\neq 0,
\ee
where $\lambda_0$ is the initial external parameter. Therefore, it is a non-degenerate case. Such an example presents $\Psi_0(0)=1/(2\beta\lambda_0^2)$~\cite{naze2022optimal}. Relation~\eqref{eq:condlrt} reads
\be
\delta\lambda \ll \sqrt{2}\lambda_0,
\ee
revealing an upper bound for the driving strength. 

To compare linear response theory with the exact case to see the consistency of the typical length as a good measurement of agreement, the irreversible work from linear response theory can be written as~\cite{naze2022optimal}
\begin{widetext}
\be
W_{\rm irr}\left(\frac{\tau}{\tau_R},\frac{\delta\lambda}{\ell_0}\right) = \frac{1}{\beta}\left(\frac{\delta\lambda}{\ell_0}\right)^2\int_0^1\int_0^1 \exp\left[-\left(\frac{\tau}{\tau_R}\right)|t-u|\right]\dot{g}(t)\dot{g}(u)dtdu,
\ee
\end{widetext}
\noindent where $\tau_R$ is its relaxation time. Interestingly, the irreversible work appears as an interplay between the driving aspects of the process ($\delta\lambda$ and $\tau$), and intrinsic equilibrium aspects of the system ($\ell_0$ and $\tau_R$), which is factored in ratios $\delta\lambda/\ell_0$ and $\tau/\tau_R$ from the double integral and the kernel argument, respectively. 

\begin{figure}[t]
    \centering
    \includegraphics[width=1\linewidth]{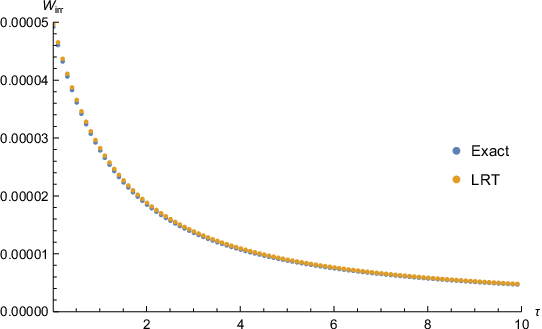}
    \caption{Matching of irreversible work curves for perturbation $\delta\lambda=\sqrt{2}\lambda_0/100$ for exact and linear response theory of the stiffening trap. It was used $\lambda_0=1$, $\beta=1$, and $\tau_R=1/2$ for the linear protocol $g(t)=t$.}
    \label{fig:1}
\end{figure}

\begin{figure}[t]
    \centering
        \includegraphics[width=1\linewidth]{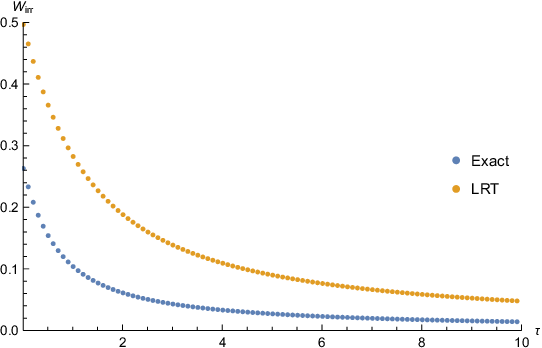}
    \caption{Disagreement of irreversible work curves for exact and linear response theory for perturbation $\delta\lambda=\sqrt{2}\lambda_0$ for the stiffening trap. It was used $\lambda_0=1$, $\beta=1$, and $\tau_R=1/2$ for the linear protocol $g(t)=t$.}
    \label{fig:2}
\end{figure}

Using such a functional and numerically solving the irreversible work without perturbation (see Ref.~\cite{pnaze_TURWD_2024}), in Fig.~\ref{fig:1}, I depict the matching of graphics of the switching time versus irreversible work for perturbation $\delta\lambda=\sqrt{2}\lambda_0/100$. It was used $\lambda_0=1$, $\beta=1$ and $\tau_R=1/2$ for the linear protocol $g(t)=t$. In Fig.~\ref{fig:2}, I depict the disagreement of graphics for exact and linear response theory for perturbation $\delta\lambda=\sqrt{2}\lambda_0$. It was used $\lambda_0=1$, $\beta=1$, and $\tau_R=1/2$ for the linear protocol $g(t)=t$ as well. Indeed, the typical length provides a quantitative diagnostic of the agreement of the irreversible work between the exact and linear response theory.

Although illustrated here for overdamped harmonic systems, the criterion itself depends only on equilibrium fluctuations through 
$\Psi_0(0)$. No assumption about linear dynamics or Gaussian statistics is required in its derivation. The harmonic examples merely provide transparent realizations where the agreement between exact and linear-response expressions can be explicitly verified. Other complex systems, like the ones in viscoelastic baths~\cite{naze2025viscoelastic}, are liable of testing.

\subsection{Kibble-Zurek mechanism in open systems: failure of linear response theory}

An interesting limit arises near Kibble-Zurek mechanism in open systems, where the equilibrium susceptibility diverges and, correspondingly, $\Psi_0(0)\rightarrow\infty$~\cite{naze2026unifying} (see the underdamped Brownian motion subject to a stiffening trap as a typical example). Near criticality, $\ell_0$ vanishes, implying that the admissible driving strength shrinks to zero. Within the present framework, this signals that linear response theory becomes increasingly fragile close to criticality: even arbitrarily small perturbations can induce corrections comparable to equilibrium fluctuations. The result is consistent with the well-known enhancement of susceptibility and critical fluctuations near continuous phase transitions, and provides a quantitative way to understand why the regime of controlled linear response collapses as criticality is approached. In real scenarios, where criticality is not fully achieved, a possible range of validity for linear response theory can still be found~\cite{naze2026unifying}.

\section{Fisher information geometry}

Another illuminating perspective arises from information geometry.
The equilibrium family of probability distributions $\rho_0(\lambda)$ defines a statistical manifold parametrized by $\lambda$. On this manifold, the Fisher information
\be
\mathcal{I}(\lambda) := \langle (\partial_\lambda \ln{\rho_0})^2\rangle_0
\ee
plays the role of a Riemannian metric tensor,
$g_{\lambda\lambda} = \mathcal{I}(\lambda)$. This metric quantifies how rapidly the equilibrium distribution changes under an infinitesimal variation of the control parameter. For nearby parameters, the relative entropy admits the quadratic expansion
\be
D_{\rm KL}(\rho_0(\lambda_0+\delta\lambda)||\rho_0(\lambda_0))\approx\frac{1}{2}\mathcal{I}(\lambda_0)\delta\lambda^2.
\ee
Thus, the Fisher information determines the local statistical distinguishability between equilibrium states. Rewriting the Fisher information in terms of the relaxation function, one has $\mathcal{I}(\lambda_0) = \beta\Psi_0(0)$, and the relation becomes
\be
\delta\lambda \ll \frac{1}{\sqrt{\mathcal{I}(\lambda_0)}}:=\ell_0.
\ee
In this manner, linear response theory functioning is strictly related to Fisher information geometry. Indeed, the typical length is the local radius of convergence of distinguishability in the equilibrium manifold. This interpretation clarifies the geometric meaning of weak driving: the perturbation must not move the system too far, in the information-geometric sense, from its initial equilibrium state.

It is important to emphasize that this local bound differs from the notion of thermodynamic length along a finite protocol, which involves integrating the metric along a trajectory in parameter space~\cite{SivakCrooks2012,ZulkowskiSivakCrooksDeWeese2012}. While thermodynamic length controls dissipation accumulated along a path, the present result identifies a local amplitude scale that bounds the perturbation itself. In this sense, $\ell_0$ separates intrinsic equilibrium geometry from protocol-dependent dynamical effects.

Also, the result derived here is not merely the consequence of a quadratic Taylor expansion of the relative entropy. While such an expansion describes local distinguishability between nearby equilibrium states, it does not by itself provide a physically meaningful bound on the admissible driving strength. In contrast, by combining the first-order structure of the probability correction with a Cauchy–Schwarz inequality and the thermodynamic interpretation of relative entropy, we obtain a self-consistent constraint that links the perturbation amplitude directly to equilibrium fluctuations. The typical length therefore emerges as an intrinsic scale set by the system’s equilibrium geometry, rather than as a truncation artifact of a series expansion.

\section{Final remarks}

In this letter, I addressed a question that is often assumed rather than examined: what precisely defines the regime of validity of linear response theory? Instead of relying on heuristic small-parameter arguments, we derived a self-consistent criterion directly from the thermodynamic uncertainty relation. The result is simple but powerful: the admissible driving strength is bounded by a typical length scale determined entirely by equilibrium fluctuations.

This typical length, expressed through the relaxation function or, equivalently, the Fisher information, reveals that the notion of ``weak driving'' is not arbitrary. It is encoded in the equilibrium structure of the system and its coupling to the environment. Linear response theory is valid when the perturbation remains small compared to this intrinsic scale, not merely compared to the initial value of the control parameter.

An appealing aspect of the result is its universality. The bound does not depend on the details of the driving protocol. In open systems, the relative entropy connects directly to heat absorption, giving the inequality a clear thermodynamic meaning. From an information-geometric perspective, the same condition emerges as a requirement that the statistical distance traveled in parameter space remains small. In this sense, thermodynamics and information theory converge naturally in characterizing the weak-driving regime.

The examples of overdamped Brownian motion in harmonic traps illustrate how the bound reproduces and clarifies conditions that are frequently used in practice. What is often imposed as a technical assumption now acquires a structural origin: it is the tightness of the FRI that governs the regime where linear response remains reliable. The brief discussion of open systems presenting the Kibble-Zurek mechanism illustrates as well how linear response theory breaks down at criticality.

More broadly, this work suggests that the validity of approximate theories can sometimes be inferred not by calculating higher-order corrections explicitly, but by examining the fundamental inequalities that structure nonequilibrium thermodynamics. The typical length identified here may serve as a useful diagnostic tool in experiments and simulations, especially in systems where the separation between weak and strong driving is not immediately obvious.

A natural next step is to explore how this idea carries over to quantum systems. In the quantum regime, fluctuations are not only statistical but also genuinely quantum, shaped by coherence and the non-commutativity of observables. This raises an interesting question: can we define a similar typical length that signals the validity of linear response when quantum effects are present? Addressing this would likely involve quantum versions of the FRI and a careful treatment of how work and measurements are defined. Beyond its technical interest, such an extension could offer deeper insight into how quantum fluctuations, information, and thermodynamics come together to set the boundaries of linear response theory.

\section*{Data availability}

The code used in this work is available in Ref.~\cite{pnaze_TURWD_2024}.

\begin{acknowledgements}
I thank Marcus V. S. Bonan\c{c}a for enlightening discussions.
\end{acknowledgements}

\bibliography{FRIWD}

\onecolumngrid

\appendix

\section{Relative entropy under linear response theory}
\label{app:A}

Consider $\rho(\Gamma,t)$ as the non-equilibrium probability distribution of the particle, and $\rho(\Gamma,0)=\rho_0(\Gamma)$ as the initial canonical distribution. The relative entropy of $\rho(\Gamma,t)$ in respect to $\rho_0(\Gamma)$ (called as well divergence of Kullback-Leibler) is defined as
\be
D_{\rm KL}(\rho(t)||\rho_0):=\int_{\Gamma} \log{\frac{\rho(\Gamma',t)}{\rho_0(\Gamma')}}\rho(\Gamma',t)d\Gamma'
\ee
Considering the expansion due to the weak driving
\be
\rho(\Gamma,t) \approx \rho_0(\Gamma)+\rho_{1}(\Gamma,t)\delta\lambda+\rho_{2}(\Gamma,t)\delta\lambda^2
\ee
The relative entropy up to second-order in $\delta\lambda$ becomes
\be
D_{\rm KL}(\rho(t)||\rho_0)\approx\frac{\delta\lambda^2}{2} \int_{\Gamma}\frac{\rho_1^2(\Gamma',t)}{\rho_0(\Gamma')}d\Gamma',
\ee
which is a positive quantity. Writing in terms of the relaxation function~\eqref{eq:relaxfunc}
\be
D_{\rm KL}(\rho(t)||\rho_0)=-\frac{\delta\lambda^2\beta}{2}\int_0^t \int_0^t \ddot{\Psi}_0(u-u')g(u)g(u')dudu',
\ee
which is, for the special case of open classical systems, an expression proportional to the heat absorbed by the system during the driving~\cite{naze2025thermodynamic}
\be
D_{\rm KL}(\rho(t)||\rho_0)=\beta Q(t).
\ee

\section{Fluctuation-response inequality}
\label{app:B}

Consider $B(\Gamma)$ as an observable. According to Cauchy-Schwarz inequality, it holds
\begin{align}
(\langle B\rangle_t-\langle B\rangle_0)^2&=\left(\int_{\Gamma} B(\Gamma')\rho_1(\Gamma',t)d\Gamma' \right)^2\\
&=\left(\int_{\Gamma} (B(\Gamma')-\langle B\rangle_0)\rho_1(\Gamma',t)d\Gamma' \right)^2\\
&=\delta\lambda^2\left(\int_{\Gamma} (B(\Gamma')-\langle B\rangle_0)\rho_1(\Gamma',t)\frac{\rho_0(\Gamma')^{1/2}}{\rho_0(\Gamma')^{1/2}}d\Gamma' \right)^2\\
&\le \left(\delta\lambda^2\int_{\Gamma} \frac{\rho_1^2(\Gamma',t)}{\rho_0(\Gamma')}d\Gamma'\right)\left(\int_{\Gamma} (B(\Gamma')-\langle B\rangle_0)^2\rho_0(\Gamma')d\Gamma'\right)\\
& = \left(\delta\lambda^2\int_{\Gamma} \frac{\rho_1^2(\Gamma',t)}{\rho_0(\Gamma')}d\Gamma'\right) (\langle B^2\rangle_0-\langle B\rangle_0^2).
\end{align}
Rearranging, one has the fluctuation-response inequality~\eqref{eq:tur0}.

\section{Quasistatic limit of relative entropy}
\label{app:c}

Using the First Law of Thermodynamics~\cite{naze2025thermodynamic}, the relative entropy can be written as
\be
D_{\rm KL}(\rho(t)||\rho_0)=\frac{\delta\lambda^2}{2}\beta \Psi_0(0)+\delta\lambda^2\beta\int_0^t \Psi_0(t-u)\dot{g}(u)du+\frac{\delta\lambda^2\beta}{2}\int_0^t \int_0^t \Psi_0(u-u')\dot{g}(u)\dot{g}(u')dudu',
\ee
Taking the limit $t\to \infty$, and knowing that the integrals go to zero due to the sufficiently fast decaying of the relaxation function, I recover the result~\eqref{eq:heatqs}.

\end{document}